\documentclass[aps,floats]{revtex4}
\usepackage{tabularx,graphicx}
\usepackage{epsfig}
\usepackage{amsmath}
\usepackage{bbm}
\usepackage{graphicx}

\begin{document}


\title{Is graphene on the edge of being a topological insulator?}

\author{J. Gonz\'alez}

\affiliation{Instituto de Estructura de la Materia,\\
        Consejo Superior de Investigaciones Cient\'{\i}ficas,\\ Serrano 123,
        28006 Madrid, Spain}

\date{\today}

\begin{abstract}
We show that, at sufficiently large strength of the long-range Coulomb 
interaction, a mass term breaking parity (so-called Haldane mass) is 
dynamically generated in the many-body theory of Dirac fermions describing 
the graphene layer. While the tendency towards chiral symmetry breaking is 
stronger than for the dynamical breakdown of parity at spatial dimension 
$D > 2$, we find that the situation is reversed at $D = 2$. The need to 
regularize the many-body theory in a gauge-invariant manner (taking the limit 
$D = 2 - \epsilon $) is what leads to the dominance of the parity-breaking 
pattern in graphene. We compute the critical coupling for the generation of 
a parity-breaking mass from the finite radius of 
convergence of the ladder series supplemented with electron self-energy 
corrections, finding a value quite close to the effective interaction strength 
for graphene in vacuum after including Fermi velocity renormalization and 
static RPA screening of the Coulomb interaction.

\end{abstract}

\maketitle



\section{Introduction}

During the last years, a great effort has been devoted to the investigation 
of the so-called graphene, the material made of a one-atom-thick carbon 
layer\cite{novo}. Many of the interesting features of graphene arise from the 
unconventional band structure of the carbon sheet, where electron 
quasiparticles behave like massless Dirac fermions\cite{geim,kim}. The 
relativistic-like invariance of the low-energy electron system is at the 
origin of outstanding transport properties\cite{rmp}, like the reduced 
influence of scatterers with size above the C-C distance\cite{ando} or the 
perfect transmission through one-dimensional potential barriers\cite{kats}.

In the graphene layer, the Coulomb repulsion between electrons constitutes 
the dominant interaction. This makes the electron system to be at low 
energies a variant of Quantum Electrodynamics, but placed here in the strong 
coupling regime as the ratio of $e^2$ to the Fermi velocity $v_F$ of the 
electrons is nominally larger than one. 
There have been already several proposals to observe unconventional signatures 
of the interacting electrons in graphene, including the anomalous screening of 
impurities carrying a sufficiently large charge\cite{nil,fog,shy,ter}. 
It has been also found that the own $e$-$e$ interaction in the 
layer should lead to a linear dependence on energy of the quasiparticle decay 
rate\cite{qlt}, as a consequence of the vanishing density of states at the 
charge neutrality point, and in agreement with measurements carried out in 
graphite\cite{exp}. 

Yet the effects of electron correlations have been quite elusive in graphene
(setting aside the observation of the fractional quantum Hall effect). In this
respect, a number of theoretical works have studied the dynamical breakdown of
the chiral symmetry of the Dirac 
fermions\cite{khves,gus,vafek,khves2,her,jur,drut1,drut2,hands,hands2,gama,fer,ggg,me,prb},
being remarkable that most part of the approaches have asserted its viability,
with a critical coupling below the nominal interaction strength of graphene in vacuum 
($e^2/4\pi v_F \approx 2.2$). Experimentally, it has been observed instead the 
increase of the Fermi velocity when approaching the charge neutrality 
point\cite{exp2}, reflecting the renormalization of the effective interaction 
strength towards a vanishing value in the low-energy theory\cite{np2,prbr}. 

It has been proposed that this scaling of the electron system approaching the 
noninteracting limit could explain the absence of any effect of dynamical gap 
generation in graphene. Some theoretical studies have indeed incorporated the 
electron self-energy corrections in the many-body theory to show that they can 
push the critical coupling for exciton condensation above the largest possible 
value attained for graphene in vacuum\cite{sabio,prb}. Anyhow, the 
theoretical analyses have dealt mainly with the static RPA screening of the 
Coulomb potential, that tends to underestimate the effective strength of the 
interaction at low energies. The consideration of a more sensible dynamical 
screening has shown to lead to smaller values of the critical coupling for the 
excitonic instability, below the value $e^2/4\pi v_F = 2.2$ \cite{ggg,prb}.      

These preceding investigations have focused on the dynamical generation 
of a gap from the formation of a staggered charge density in the graphene
lattice, while the possible condensation of other order parameters has been
mostly overlooked (see however Refs. \cite{rag,herbut}). Nevertheless, already 
in the early studies of Quantum 
Electrodynamics at spatial dimension $D = 2$, it was realized that the
dynamical symmetry breaking could develop in two different ways. That is, 
one could have the condensation of a mass breaking the chiral symmetry but 
preserving the invariance under parity of the Dirac fermions\cite{pis,qed,kog}, 
or otherwise a mass maintaining the chiral symmetry but breaking the 
invariance under parity\cite{parb,sem1}. The prevalence of one pattern of 
symmetry breaking over the other was shown to depend in general on the 
balance between different local four-fermion interactions in the electron 
system\cite{sem1}.

The two different order parameters characterizing the dynamical symmetry
breaking, the parity-invariant mass and the parity-breaking mass, are not 
related by any fundamental symmetry in the graphene electron system, which 
means that they can be generated quite independently. The possibility of 
introducing a mass operator breaking the invariance under parity was 
considered in the context of a tight-binding model of electrons in the 
honeycomb lattice in the seminal work of Ref. \cite{haldane}. The dynamical 
generation of the parity-breaking mass establishes in fact a bridge between 
graphene and the so-called topological insulators, as one of the consequences 
of the condensation of the parity-breaking order parameter would be the 
spontaneous development of loop currents in the honeycomb lattice.

In this work we study the dynamical generation of the parity-breaking mass 
term in the theory of Dirac fermions in graphene, taking the long-range 
Coulomb repulsion as the relevant interaction between electrons. We will 
consider the dynamical symmetry breaking analyzing the corresponding order 
parameter in the many-body theory, comparing it with the more conventional 
effect of exciton condensation breaking chiral symmetry but preserving 
parity. We recall that the space of Dirac fermions at $D = 2$ is defined by 
the algebra of matrices $\{ \gamma_{\sigma} \}$ ($\sigma = 0,1,2$) satisfying 
the anticommutation relations 
\begin{equation}
\{ \gamma_\mu, \gamma_\nu \} = 2 \: {\rm diag } (1,-1,-1)
\label{ac}
\end{equation}
In this notation, the order parameter for the parity-invariant mass will 
correspond to the fermion density averaged with the $\gamma_0 $ matrix, while 
the generation of the parity-breaking mass will be given by a nonvanishing 
average of the fermion current $i \epsilon_{3ij} \gamma_i \gamma_j $.

One can check that, at $D = 2$, the product of all the
$\gamma $ matrices, $\gamma_5 \equiv \gamma_0 \gamma_1 \gamma_2 $, commutes 
with any other matrix in the space of the Dirac fermions, so that its 
action is equivalent to the identity. This means that, computing for instance 
the ladder series of many-body corrections, all contributions to the $\gamma_0$ 
vertex should be formally the same as those to the vertex for the current 
$i \epsilon_{3ij} \gamma_i \gamma_j $. The sum of the ladder series provides 
a particularly sensible approach to dynamical symmetry breaking, as it encodes 
the most divergent diagrams at each level of the perturbative 
expansion\cite{mis}. It is actually the need to devise a consistent 
regularization of those divergences which plays a central role in the 
evaluation of the different order parameters. 

We will show that, in the 
theory regularized with a method chosen specifically to preserve its gauge 
invariance (such as dimensional regularization), the perfect match between the 
order parameters for the parity-breaking and the parity-invariant mass is lost. 
This situation is similar to the well-known case of conventional Quantum 
Electrodynamics at spatial dimension $D = 3$, where there is no regularization 
that can simultaneously preserve the gauge and the chiral symmetries, which 
makes unavoidable to discard the latter in favor of the first.

In the case of graphene, the use of dimensional regularization has shown to be 
a most suitable choice to preserve the underlying gauge invariance of the 
many-body theory\cite{np2,juricic,jhep}. Recently, it has been checked that 
such a method is singled out as the right approach to reproduce the corrections 
to the DC conductivity calculated from the lattice model\cite{ros}.
In our problem, the need to regularize the many-body theory while preserving 
gauge invariance is also what produces the anomalous mismatch between the 
respective vertices for $\gamma_0$ and the current 
$i \epsilon_{3ij} \gamma_i \gamma_j $. 

We will see that, at $D$ greater than 
2, the divergent corrections to the vertex for the parity-invariant mass are 
in general greater that those to the vertex for the parity-breaking mass, 
while the situation is reversed for $D < 2$. The diagrams in the Dirac 
many-body theory are made convergent by analytic continuation to spatial 
dimension $D = 2 - \epsilon$, which explains that the tendency to develop a 
nonvanishing order parameter for the parity-breaking mass becomes dominant in
the physical renormalized theory at $D \rightarrow 2$. As a 
consequence, we will see that there is a critical coupling for the dynamical 
breakdown of parity in graphene which turns out to be smaller than the 
critical coupling obtained in the more conventional mechanism relying on 
chiral symmetry breaking.

In the next section, we set up the framework to describe the many-body theory
of Dirac fermions, highlighting the classical scale invariance that opens the
way to apply the renormalization group approach. We focus on the computation 
of the vertices for $\gamma_0$ and the current $i \epsilon_{3ij} \gamma_i \gamma_j $
in Sec. III, where we show that it is possible to define cutoff-independent 
observable quantities from the sum of ladder diagrams. We find that 
the two ladder series for the renormalized vertices have finite radii of 
convergence, that we compute in each case to obtain the respective critical 
couplings signaling the dynamical generation of the parity-invariant and the 
parity-breaking mass. Finally, we discuss in Sec. IV the physical implications 
of the dynamical symmetry breaking arising from a parity-breaking mass.

\section{Dirac many-body theory}

The low-energy electron quasiparticles are disposed in graphene into conical 
conduction and valence bands that touch at the six corners of the Brillouin 
zone\cite{rmp}. There are two inequivalent classes of electronic states, that
can be disposed around a pair of independent Fermi points. Thus, the low-energy 
electronic excitations can be encoded into a couple of four-dimensional Dirac 
spinors $\{ \psi_i \}$,  which are characterized by having linear energy-momentum 
dispersion $\varepsilon ({\bf p}) = v_F |{\bf p}|$. The index $i$ accounts here 
for the two independent spin projections, but it will be omitted to simplify the
notation in what follows. The kinetic term of the hamiltonian 
in this low-energy theory is given by 
\begin{equation}
H_0 = - i v_F \int d^2 r \; \overline{\psi}({\bf r}) 
 \boldsymbol{\gamma}   \cdot \boldsymbol{\nabla}  \psi ({\bf r}) 
\label{h0} 
\end{equation}
where $\overline{\psi} = \psi^{\dagger} \gamma_0 $ and $\{ \gamma_{\sigma} \}$
is a collection of four-dimensional matrices satisfying the relations (\ref{ac}). 
It is convenient to represent them in terms of Pauli matrices as
\begin{equation}
\gamma_{0,1,2} = (\sigma_3, \sigma_3 \sigma_1, \sigma_3 \sigma_2) \otimes
 \tau_3
\end{equation} 
where the first factor acts on the two sublattice components of 
the graphene honeycomb lattice and the second factor operates on the set of 
two independent Fermi points.

The dominant $e$-$e$ interaction is given in graphene by the long-range 
Coulomb repulsion. This interaction does not have a conventional screening at 
long distances, given the vanishing density of states at the Fermi 
points connecting the conduction and valence bands. A suitable description
of the Coulomb interaction starts then by considering the unscreened potential 
$V({\bf r}) = e^2/4\pi |{\bf r}|$. The full hamiltonian including the $e$-$e$ 
interaction can be represented as 
\begin{eqnarray}
H   =    - i v_F \int d^2 r \; \overline{\psi}({\bf r}) 
 \mbox{\boldmath $\gamma   \cdot \nabla $} \psi ({\bf r}) 
         + \frac{e^2}{8 \pi} \int d^2 r_1
\int d^2 r_2 \; \rho ({\bf r}_1) 
       \frac{1}{|{\bf r}_1 - {\bf r}_2|} \rho ({\bf r}_2)  \;\;\;\;\;
\label{ham}
\end{eqnarray}
with $\rho ({\bf r}) = \overline{\psi} ({\bf r}) \gamma_0 \psi ({\bf r})$.

The prevalence of the Coulomb interaction at low energies can be understood 
from the scaling properties of the system governed by (\ref{ham}). The 
long-range Coulomb repulsion is the only interaction that is not suppressed, 
at the classical level, when scaling the many-body theory towards the limit of 
very large distances or very low energies. From (\ref{ham}), the total action 
of the system is given by
\begin{eqnarray}
S  & = &    \int dt  \int d^2 r \; \overline{\psi}({\bf r}) (i\gamma_0 \partial_t  
 + i v_F \mbox{\boldmath $\gamma   \cdot \nabla $} ) \psi ({\bf r})    \nonumber \\
  &   &     - \frac{e^2}{8 \pi} \int dt  \int d^2 r_1
\int d^2 r_2 \; \rho ({\bf r}_1) 
       \frac{1}{|{\bf r}_1 - {\bf r}_2|} \rho ({\bf r}_2)  \;\;\;\;\;
\label{s}
\end{eqnarray}
This action is invariant under the combined 
transformation of the space and time variables and the scale of the fields
\begin{equation}
t' = s t  \;\; , \;\;  {\bf r}' = s {\bf r}  \;\; , \;\; \psi' = s^{-1} \psi 
\label{scale}
\end{equation}
This means in particular that the strength of the Coulomb interaction is not 
diminished when zooming into the low-energy limit $s \rightarrow \infty $.
One can check that any other $e$-$e$ interaction without the $1/|{\bf r}|$ 
tail, as those that arise effectively from phonon exchange, would be 
suppressed at least by a power of $1/s$ under the change of variables 
(\ref{scale}), implying its irrelevance in the low-energy limit.

On the other hand, the many-body theory does not preserve in general the scale 
invariance of the classical action (\ref{s}), as a high-energy cutoff 
has to be introduced to obtain finite results in the computation of 
many-body corrections to different observables. The analysis of the cutoff 
dependence of the many-body theory provides deeper insight into the effective 
low-energy theory. If the theory is renormalizable, it must be possible to 
absorb all powers of the cutoff dependence into a redefinition of the 
parameters in the action (\ref{s}). This should be therefore modified to read
\begin{eqnarray}
\lefteqn{S =  Z_{\rm kin } \int dt  \int d^2 r \; \overline{\psi}({\bf r}) 
  (i\gamma_0 \partial_t  + i  Z_v \: v_F  \boldsymbol{\gamma}   
                        \cdot \boldsymbol{\nabla} ) \psi ({\bf r}) }   \nonumber \\
  &   &     - Z_{\rm int } \frac{e^2}{8 \pi} \int dt  \int d^2 r_1
\int d^2 r_2 \; \rho ({\bf r}_1) 
       \frac{1}{|{\bf r}_1 - {\bf r}_2|} \rho ({\bf r}_2)  \;\;\;\;\;
\label{sren}
\end{eqnarray}
The assumption is that $Z_{\rm kin }, Z_v$ and  $Z_{\rm int }$ (and other 
renormalization factors for composite operators that do not appear in 
(\ref{s})) can only depend on the cutoff, while they must be precisely chosen 
to render all electronic correlators cutoff independent\cite{amit}.

This possibility of absorbing all the divergent dependences on the cutoff in
a finite number of renormalization factors has been checked in the above theory 
in the limit of a large number $N$ of Dirac fermion flavors\cite{prbr,ale}, as 
well as in the ladder approximation to vertex diagrams supplemented by electron 
self-energy corrections\cite{jhep}. We will use this latter approach in the 
computation of the 
vertices for $\gamma_0$ and the current $i \epsilon_{3ij} \gamma_i \gamma_j $.
We recall that in the ladder approximation there is no need to renormalize
the electron quasiparticle weight, so that $Z_{\rm kin } = 1$, while the 
renormalization factor $Z_v$ for the Fermi velocity coincides with the 
expression obtained at the first perturbative level\cite{jhep}. The 
computation of the expectation value of other fermion currents requires 
however the introduction of new renormalization factors in order to obtain 
finite, cutoff-independent results, as shown in Ref. \cite{jhep}. These 
renormalization factors encode in general valuable information about the 
possible condensation of different excitations and the corresponding dynamical 
symmetry breaking.

The gapless character of the electronic spectrum is in principle protected 
by two different symmetries of the hamiltonian (\ref{ham}). This is invariant 
under a $Z_2$ chiral symmetry ($CS$) that consists in the exchange of the 
two-dimensional spinors attached to the two independent Fermi points
\begin{equation}
CS:\psi ({\bf r})  \rightarrow  \left( 1 \otimes  \tau_1 \right) \psi ({\bf r}) 
\label{cs}
\end{equation}
We could add to the hamiltonian (\ref{ham}) a term breaking the invariance 
under (\ref{cs})
\begin{equation}
H_m   =  m \int d^2 r \; \overline{\psi}({\bf r})  \psi ({\bf r}) 
\label{pim}
\end{equation}
while respecting the space-time symmetries\cite{sem}. This term, 
corresponding to a conventional parity-invariant mass, would certainly open a 
gap in the electronic spectrum. But this same effect can be obtained with a 
different perturbation preserving the chiral symmetry (\ref{cs})
\begin{equation}
H_{m'}   =  m' \int d^2 r \; \overline{\psi}({\bf r}) 
    \;  i \gamma_0 \: \epsilon_{3ij} \gamma_i \gamma_j \; \psi ({\bf r}) 
\label{pbm}
\end{equation}
One can check that (\ref{pbm}) breaks instead the invariance under parity ($P$)
\begin{equation}
P:\psi (x,y)  \rightarrow  \left( \sigma_1 \otimes  \tau_1 \right) \psi (x,-y) 
\label{par}
\end{equation}
The term (\ref{pbm}) corresponds to the parity-breaking mass mentioned 
before\cite{haldane}. It can be shown that there are no other operators (not 
implying the electron spin or the exchange
of excitations about the two Fermi points) that may open a gap in the 
electronic spectrum\cite{cham,mas}. We finally recall that, while (\ref{pim}) or 
(\ref{pbm}) may not be present nominally in the hamiltonian, they can be 
generated in the many-body theory at sufficiently large interaction strength, 
leading to the dynamical breakdown of the chiral symmetry or parity in each 
respective case.

\section{Renormalization of staggered charge density and loop current operators}

We focus then on the possible development of a vacuum expectation value 
for the parity invariant operator
\begin{equation}
\rho_3 ({\bf r}) = \overline{\psi} ({\bf r})  \psi ({\bf r})
\end{equation}
as well as for the parity-breaking bilinear
\begin{equation}
\widetilde{\rho}_3 ({\bf r}) 
            = \overline{\psi} ({\bf r}) \gamma_0 \Omega_3 \psi ({\bf r})
\end{equation}
with 
\begin{equation}
\Omega_3 =  i \epsilon_{3ij}  \gamma_i \gamma_j
\end{equation}
We will characterize such a phenomenon by inspection of the respective 
vertices
\begin{equation}
 \Gamma_3 ({\bf q},\omega_q;{\bf k},\omega_k)  =
   \langle  \rho_3 ({\bf q},\omega_q) 
        \psi ({\bf k}+{\bf q},\omega_k + \omega_q) 
           \overline{\psi} ({\bf k},\omega_k) \rangle_{\rm 1PI}
\label{veven}
\end{equation}
and 
\begin{equation}
 \widetilde{\Gamma}_3 ({\bf q},\omega_q;{\bf k},\omega_k)  =
   \langle  \widetilde{\rho}_3 ({\bf q},\omega_q) 
        \psi ({\bf k}+{\bf q},\omega_k + \omega_q) 
           \overline{\psi} ({\bf k},\omega_k) \rangle_{\rm 1PI}
\label{vodd}
\end{equation}
where 1PI denotes that we take the one-particle irreducible part of the
correlator. We are going to see that the above vertices have a finite radius
of convergence in perturbation theory, which is the signature of the dynamical
breakdown of symmetry driven by the interaction.

\begin{figure}

\vspace{0.5cm}

\begin{center}
\mbox{\epsfxsize 10cm \epsfbox{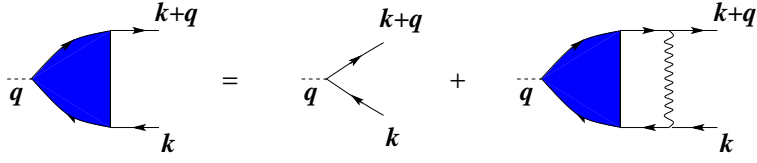}} 

\end{center}
\caption{Self-consistent diagrammatic equation for a generic vertex $\Gamma_i$
in the ladder approximation.}
\label{one}
\end{figure}

A sensible approach to determine the critical interaction strengths at 
which (\ref{veven}) and (\ref{vodd}) diverge consists in performing the sum 
of ladder diagrams for the two vertices. The ladder contributions contain 
indeed the most divergent part of the vertex at each level of the perturbative 
expansion\cite{mis}. The ladder series built in that way can be easily encoded 
in the self-consistent diagrammatic equation represented in Fig. \ref{one}.  
In what follows, we will see how to obtain the respective critical couplings 
in a manner that they do not depend on the high-energy cutoff needed to 
regularize the theory, relying only on observable quantities of the 
electron system.

\subsection{Loop current vertex}

We study first the vertex for the loop current operator. In order to 
make the approach covariant and valid in any spatial dimension, we
consider the correlations of the loop operator
\begin{equation}
\widetilde{\rho}_{ij} ({\bf r}) 
         =  \overline{\psi} ({\bf r}) \gamma_0 
           i [ \gamma_i , \gamma_j ] \psi ({\bf r})
\end{equation}
This is now a second-rank tensor, leading to the vertex
\begin{equation}
 \widetilde{\Gamma}_{ij} ({\bf q},\omega_q;{\bf k},\omega_k)  =
   \langle  \widetilde{\rho}_{ij} ({\bf q},\omega_q) 
        \psi ({\bf k}+{\bf q},\omega_k + \omega_q) 
           \overline{\psi} ({\bf k},\omega_k) \rangle_{\rm 1PI}
\label{vcov}
\end{equation}

In the ladder approximation, the vertex 
$\widetilde{\Gamma}_{ij} ({\bf q},\omega_q;{\bf k},\omega_k)$
has to satisfy in the limit ${\bf q} = 0$ and $\omega_q = 0$ the self-consistent
equation
\begin{eqnarray}
\lefteqn{\widetilde{\Gamma}_{ij}  ({\bf 0},0;{\bf k},\omega_k) = }
                                                    \nonumber       \\
  &  &   i \gamma_0  [ \gamma_i , \gamma_j ]   +   
      i \frac{e^2_0}{2} \int \frac{d^D p}{(2\pi )^D} \frac{d\omega_p}{2\pi } 
     \gamma_0 \frac{-\gamma_0 \omega_p  + v_F \boldsymbol{\gamma} \cdot {\bf p} }
                 {-\omega_p^2  + v_F^2 {\bf p}^2 - i\eta } 
    \widetilde{\Gamma}_{ij}  ({\bf 0},0;{\bf p},\omega_p) 
      \frac{-\gamma_0 \omega_p  + v_F \boldsymbol{\gamma} \cdot {\bf p} }
                 {-\omega_p^2  + v_F^2 {\bf p}^2 - i\eta } \gamma_0
               \frac{1}{|{\bf k}-{\bf p}|}   \;\;\;\;\;\;\;
\label{selfh}
\end{eqnarray}
where we have regularized the momentum integrals by continuing the 
number of spatial dimensions to $D = 2 - \epsilon$. Accordingly, $e_0$ 
is now a dimensionful coupling which demands the introduction of an 
auxiliary momentum scale $\mu $ through the relation
\begin{equation}
e_0 = \mu^{\epsilon /2}  e
\end{equation}

At this point we may resort to a perturbative resolution for the
vertex, noticing that it can only depend on the dimensionless ratio
$\mu / |{\bf k}|$. 
It is easily realized that the operator with the tensor 
$[ \gamma_i , \gamma_j ]$ can mix with the other antisymmetric
second-rank tensor structure one can build, corresponding to 
$\gamma_i k_j - k_i \gamma_j $. Thus, the most general solution 
of Eq. (\ref{selfh}) can be expressed as a power series in the coupling 
$\lambda_0 = e_0^2/4\pi v_F $
\begin{equation}
\widetilde{\Gamma}_{ij}   ({\bf 0},0;{\bf k},\omega_k)  = 
    i \gamma_0  [ \gamma_i , \gamma_j ]   
 \left(1 + \sum_{n=1}^{\infty} \lambda_0^n 
                           \frac{t_n }{|{\bf k}|^{n\epsilon}} \right)
   +  i \gamma_0 (\gamma_i n_j - n_i \gamma_j ) 
                    (\boldsymbol{\gamma} \cdot {\bf n})
            \sum_{n=1}^{\infty} \lambda_0^n 
                           \frac{t_n' }{|{\bf k}|^{n\epsilon}}
\label{exp}
\end{equation}
where we have called ${\bf n} = {\bf k}/|{\bf k}|$.

Each term in the series (\ref{exp}) can
be obtained by means of an iterative procedure from the previous 
orders. By inserting a term proportional to 
$\gamma_0  [ \gamma_i , \gamma_j ]$ at the right-hand-side of
Eq. (\ref{selfh}), we get for instance 
\begin{eqnarray}
 - \frac{e^2_0}{2} \int \frac{d^D p}{(2\pi )^D} \frac{d \overline{\omega}_p}{2\pi } 
     \gamma_0 \frac{-i \gamma_0 \overline{\omega}_p  + v_F \boldsymbol{\gamma} \cdot {\bf p} }
                 {\overline{\omega}_p^2  + v_F^2 {\bf p}^2  } 
        \gamma_0  [ \gamma_i , \gamma_j ]  \frac{1}{|{\bf p}|^{n\epsilon} }
      \frac{-i \gamma_0 \overline{\omega}_p  + v_F \boldsymbol{\gamma} \cdot {\bf p} }
                 {\overline{\omega}_p^2  + v_F^2 {\bf p}^2 } \gamma_0
               \frac{1}{|{\bf k}-{\bf p}|}                \nonumber  \\
   =  \lambda_0 \gamma_0  [ \gamma_i , \gamma_j ]  
                       \frac{A_n(\epsilon)}{|{\bf k}|^{(n+1)\epsilon} }
    + \lambda_0 \gamma_0 (\gamma_i n_j - n_i \gamma_j )  (\boldsymbol{\gamma} \cdot {\bf n})
                       \frac{B_n(\epsilon)}{|{\bf k}|^{(n+1)\epsilon} }  
\end{eqnarray}
with 
\begin{eqnarray}
A_n(\epsilon) & = & \frac{(4\pi)^{\epsilon/2}}{4} 
     \frac{\Gamma \left(\tfrac{n+1}{2}\epsilon \right) 
           \Gamma \left(\tfrac{1-(n+1)\epsilon}{2} \right) 
           \Gamma \left(\tfrac{3-\epsilon}{2} \right)}
               {\sqrt{\pi} \Gamma \left(\tfrac{3+n\epsilon}{2} \right) 
            \Gamma \left(2-\tfrac{(n+2)\epsilon}{2} \right) }          \\
B_n(\epsilon) & = & \frac{(4\pi)^{\epsilon/2}}{2} 
     \frac{\Gamma \left(1 + \tfrac{n+1}{2}\epsilon \right) 
           \Gamma \left(\tfrac{3-(n+1)\epsilon}{2} \right) 
           \Gamma \left(\tfrac{1-\epsilon}{2} \right)}
               {\sqrt{\pi} \Gamma \left(\tfrac{3+n\epsilon}{2} \right) 
                 \Gamma \left(2-\tfrac{(n+2)\epsilon}{2} \right) }
\end{eqnarray}
Moreover, when a term proportional to 
$\gamma_0 (\gamma_i n_j - n_i \gamma_j )$ is introduced 
at the right-hand-side of the integral equation, we obtain
\begin{eqnarray}
 - \frac{e^2_0}{2} \int \frac{d^D p}{(2\pi )^D} \frac{d \overline{\omega}_p}{2\pi } 
     \gamma_0 \frac{-i \gamma_0 \overline{\omega}_p  + v_F \boldsymbol{\gamma} \cdot {\bf p} }
                 {\overline{\omega}_p^2  + v_F^2 {\bf p}^2  } 
       \gamma_0 (\gamma_i n_j - n_i \gamma_j ) 
                  (\boldsymbol{\gamma} \cdot {\bf n})  \frac{1}{|{\bf p}|^{n\epsilon} }
      \frac{-i \gamma_0 \overline{\omega}_p  + v_F \boldsymbol{\gamma} \cdot {\bf p} }
                 {\overline{\omega}_p^2  + v_F^2 {\bf p}^2 } \gamma_0
               \frac{1}{|{\bf k}-{\bf p}|}            \nonumber  \\
   = \frac{1}{2} \lambda_0 \gamma_0  [ \gamma_i , \gamma_j ]  
                       \frac{A_n(\epsilon)}{|{\bf k}|^{(n+1)\epsilon} }
 + \frac{1}{2} \lambda_0 \gamma_0 (\gamma_i n_j - n_i \gamma_j )  (\boldsymbol{\gamma} \cdot {\bf n})
                       \frac{B_n(\epsilon)}{|{\bf k}|^{(n+1)\epsilon} }
\end{eqnarray}

The above formulas can be summarized in the recursion relations
\begin{eqnarray}
t_{n+1}  & = &   A_n(\epsilon)
       \left( t_n +  \frac{1}{2} t_n'   \right)      \label{re1}      \\
t_{n+1}'  & = &   B_n(\epsilon)
           \left( t_n +  \frac{1}{2} t_n'   \right)    
\label{re2}
\end{eqnarray}
In compact form, one can also write the relation
\begin{equation}
t_{n+1} +  \frac{1}{2} t_{n+1}' =   
 \left(  A_n(\epsilon) + \frac{1}{2} B_n(\epsilon) \right)
    \left( t_n +  \frac{1}{2} t_n' \right)
\label{rec}
\end{equation}

We observe that the result of computing the integral in (\ref{selfh}) 
always diverges in the limit $\epsilon \rightarrow 0$, leading to a sequence 
of higher-order poles in the $\epsilon $ parameter as we look at higher 
perturbative levels in the solution of the self-consistent equation. 
Good news are however that all these divergences can be reabsorbed after a 
multiplicative renormalization by a single factor $Z_h$, ending up with a 
finite vertex in the limit $\epsilon \rightarrow 0$
\begin{equation}
\widetilde{\Gamma}_{ij , {\rm ren}} = Z_h \widetilde{\Gamma}_{ij}  
\label{mult3}
\end{equation}
The renormalization factor $Z_h$ can have in general the structure
\begin{equation}
Z_h = 1 + \sum_{i=1}^{\infty} \frac{h_i (\lambda )}{\epsilon^i}
\label{poleh}
\end{equation}
The important point is that it can be shown that 
$\widetilde{\Gamma}_{ij , {\rm ren}} $ can be made finite at 
$\epsilon = 0$ with a set of functions $h_i (\lambda )$ that do not depend 
on the external momenta of the vertex. This is the hallmark of 
renormalizability, by which one can reabsorb the cutoff dependences into 
the renormalization factors for a few local operators of the theory.

Most interestingly, this procedure of renormalization works in the present 
case even when self-energy corrections are considered in the internal 
electron and hole states of the vertex. These supplementary contributions
to the ladder series can be taken systematically into account by replacing 
the constant $v_F$ in Eq. (\ref{selfh}) by the effective Fermi velocity
$\widetilde{v}_F({\bf p})$ found after dressing the Dirac propagator with 
the electron self-energy correction $\Sigma $
\begin{equation}
\gamma_0 \omega  - v_F \boldsymbol{\gamma} \cdot {\bf p} \; \rightarrow  \;
  \gamma_0 \omega  - v_F \boldsymbol{\gamma} \cdot {\bf p} - \Sigma 
\end{equation}
Recalling the expression of the electron self-energy in the same ladder 
approximation\cite{jhep}, we get
\begin{equation}
\widetilde{v}_F({\bf p}) = v_F + \frac{e_0^2}{16 \pi^2} 
   (4\pi )^{\epsilon /2}   
  \frac{\Gamma \left(\tfrac{1}{2}\epsilon \right) 
              \Gamma \left(\tfrac{1-\epsilon}{2}\right) 
               \Gamma \left(\tfrac{3-\epsilon}{2}\right) }
  {  \Gamma (2 - \epsilon) }
   \frac{1}{|{\bf p}|^{\epsilon}}
\label{veff}
\end{equation}
The effective $\widetilde{v}_F({\bf p})$ is in principle singular in 
the limit $\epsilon \rightarrow 0$, but the divergence can be absorbed 
by a simple renormalization of the Fermi velocity
\begin{equation}
v_F = Z_v v_{F, {\rm ren}}
\label{vren}
\end{equation}
Taking the simple pole structure 
\begin{equation}
Z_v = 1 + b_1 \frac{1}{\epsilon}
\label{zv}
\end{equation}
one needs to set $b_1 = -e^2/16\pi v_{F, {\rm ren}}$ in order
to render $\widetilde{v}_F({\bf p})$ finite at $\epsilon \rightarrow 0$ 
as a function of $v_{F, {\rm ren}}$. 

The remarkable point is that the vertex $\widetilde{\Gamma}_{ij , {\rm ren}}$, 
supplemented with the electron self-energy corrections, can be made also 
finite in the limit $\epsilon \rightarrow 0$ with a modified set of 
coefficients in (\ref{poleh}), expressed now as functions of the renormalized 
coupling
\begin{equation}
\lambda = \frac{e^2}{4\pi v_{F, {\rm ren}}}
\end{equation}
The first orders in the analytic computation of the residues 
$h_i (\lambda )$, including the effect of the electron self-energy 
corrections, turn out to be 
\begin{eqnarray}
h_1 (\lambda )  & = &   - \frac{1}{2} \lambda - \frac{1 + 4\log(2)}{32} \: \lambda^2 
      - \frac{-3+\pi ^2+12 \log (2) (3+10\log (2))}{1152}   \lambda^3     \nonumber   \\
 & &  -  \frac{9+3 \pi ^2-36 \log (2)+20 \pi ^2 \log (2)+504 \log ^2(2)+1376 \log ^3(2)+30 \zeta (3)}{12288} 
                                                     \: \lambda^4      \nonumber  \\
 & &  -  \frac{1}{737280} \left( 13 \pi ^4+6 \pi ^2 (-3+4 \log (2) (21+86 \log (2))) \right.   \nonumber   \\
 & &   \left. + 6 (51+162 \zeta (3)+4 \log (2) (45+2 \log (2) (-45+4 \log (2) (235+537 \log (2)))
                           +222 \zeta (3))) \right) 
                                   \: \lambda^5     +    \ldots    \label{first}       \\
h_2 (\lambda ) & = &  \frac{1}{16} \: \lambda^2 + 
            \frac{1 + 4\log(2)}{96}  \: \lambda^3                         
         + \frac{-6+5 \pi ^2+12 \log (2) (21+62 \log (2))}{18432}  \lambda^4          \nonumber  \\      
 & &       + \frac{\pi ^2 (37+220 \log (2))+
          3 (17+4 \log (2) (-7+2 \log (2) (299+716 \log (2)))+90 \zeta (3))}{368640 }  \: \lambda^5 
                                                               +  \ldots                     \\
h_3 (\lambda ) & = & - \frac{1 + 4\log (2)}{3072}  \: \lambda^4 
           - \frac{12+\pi ^2+156 \log (2)+360 \log ^2(2)}{184320} \lambda^5     + \ldots            \\
h_4 (\lambda ) & = & - \frac{1 + 4\log (2)}{30720}  \: \lambda^5     +  \ldots    
\label{last}
\end{eqnarray}

The function $h_1 (\lambda )$ has particular significance since it can be used
to determine the anomalous dimension of the vertex, which comes from the dependence
on the dimensionful parameter $\mu $ of the renormalized theory\cite{amit}
\begin{equation}
\widetilde{\Gamma}_{ij , {\rm ren}} \sim  \mu^{\gamma_h}
\end{equation}
The exponent $\gamma_h $ can be defined in terms of the renormalization 
factor $Z_h $ by 
\begin{equation}
\gamma_h = \frac{\mu }{Z_h}  \frac{\partial Z_h}{\partial \mu}
\label{adm}
\end{equation}
When applying Eq. (\ref{adm}), it is a highly nontrivial result that all the
pole contributions coming from (\ref{poleh}) may cancel out to give a finite
answer for $\gamma_h $ at $\epsilon \rightarrow 0$. This happens whenever
one can enforce the set of conditions\cite{ram}
\begin{equation}
\frac{d}{d\lambda } h_{i+1} (\lambda) 
  -  h_i (\lambda) \: \frac{d}{d\lambda } h_1 (\lambda)
   + b_1 (\lambda) \: \frac{d}{d\lambda } h_i (\lambda)   =   0
\label{rem}
\end{equation}
In that case, the anomalous exponent $\gamma_h $ is given by
\begin{equation}
\gamma_h = -\lambda \frac{d}{d\lambda } h_1 (\lambda)
\label{gamh}
\end{equation}

\begin{figure}

\vspace{0.5cm}

\begin{center}
\mbox{\epsfxsize 7cm \epsfbox{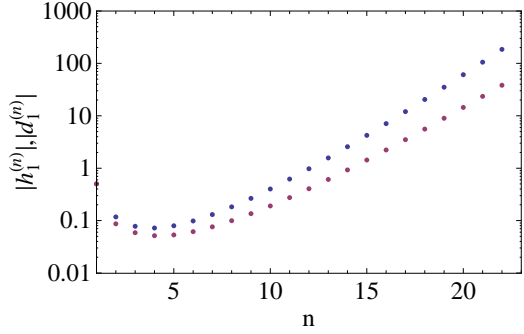}} 

\end{center}
\caption{Plot of the absolute value of the coefficients $h_1^{(n)}$ (upper 
points) and $d_1^{(n)}$ (lower points) in the respective expansions of 
$h_1 (\lambda )$ and $d_1 (\lambda )$ as power series of the renormalized 
coupling $\lambda $.}
\label{two}
\end{figure}

One can see that the analytic expressions (\ref{first})-(\ref{last}) satisfy
the conditions (\ref{rem}), and we have checked that these hold also 
for the numerical solution of the functions $h_i (\lambda )$ that we have
obtained up to order $\lambda^{22}$ in the perturbative expansion. The 
characterization of dynamical symmetry breaking from $\gamma_h$ turns out to 
be then very convenient since, according to (\ref{gamh}), the exponent is a 
scale invariant quantity depending only on the renormalized coupling $\lambda$.
It is actually observed in the numerical resolution that the perturbative 
orders of $h_1 (\lambda)$ have a geometric growth, as shown in Fig. \ref{two}. 
We can then use the same scaling technique as in Ref. \cite{jhep}, which 
allowed to reproduce there the value of the critical coupling obtained for 
chiral symmetry breaking from the resolution of the gap equation\cite{gama}.
From the expansion up to order $\lambda^{22}$, we get now the result for the 
finite radius of convergence of $\gamma_h$
\begin{equation}
\lambda_c \approx 0.508
\label{crit}
\end{equation}

One has to keep in mind that this singularity at $\lambda_c $ implies the 
divergence of the correlators of the loop current operator 
$\widetilde{\rho}_3 ({\bf r})$, as they all need to be renormalized by factors 
of $Z_h$. This implies that their anomalous dimensions are given in general by 
multiples of $\gamma_h $. We find therefore at $\lambda_c $ the onset of a new 
phase of the electron system that must correspond to dynamical symmetry
breaking, as it is characterized by the appearance of a nonvanishing 
expectation value of the operator $\widetilde{\rho}_3 ({\bf r})$ at that 
critical coupling.

\subsection{Staggered charge density vertex}

We review the computation of the vertex (\ref{veven}), which has been 
addressed in Ref. \cite{prb}, in order to establish here the comparison with the
loop current vertex. In the ladder approximation, 
$\Gamma_3 $ is proportional to the unit matrix, and the 
resolution of the equation in Fig. \ref{one} proceeds by noticing that 
\begin{align}
  \gamma_0 \frac{ -\gamma_0 \omega_p  + v_F \boldsymbol{\gamma} \cdot {\bf p} }
                      {-\omega_p^2  + v_F^2 {\bf p}^2 - i\eta } 
     \: \Gamma_3 ({\bf 0},0;{\bf p},\omega_p) \:
\frac{ -\gamma_0 \omega_p  + v_F \boldsymbol{\gamma} \cdot {\bf p} }
          {-\omega_p^2  + v_F^2 {\bf p}^2 - i\eta } \gamma_0    
  = -\frac{\Gamma_3 ({\bf 0},0;{\bf p},\omega_p)}{-\omega_p^2  + v_F^2 {\bf p}^2 - i\eta } 
\end{align}  
We arrive then at the self-consistent equation
\begin{equation}
\Gamma_3 ({\bf 0},0;{\bf k},\omega_k) = 1 -  
    i \frac{e^2_0}{2}  \int \frac{d^D p}{(2\pi )^D} \frac{d\omega_p}{2\pi } 
      \Gamma_3 ({\bf 0},0;{\bf p},\omega_p) 
          \frac{1}{-\omega_p^2  + v_F^2 {\bf p}^2 - i\eta } 
               \frac{1}{|{\bf k}-{\bf p}|}
\label{selfg3}
\end{equation}

The solution of Eq. (\ref{selfg3}) does not depend on the frequency $\omega_k$,
and the self-consistent equation becomes
\begin{equation}
\Gamma_3 ({\bf 0},0;{\bf k},\omega_k) = 1  +  \frac{e_0^2}{4} 
   \int \frac{d^D p}{(2\pi )^D} \Gamma_3 ({\bf 0},0;{\bf p},\omega_k) 
    \frac{1}{v_F |{\bf p}|} \frac{1}{|{\bf k}-{\bf p}|}
\label{selfcons}
\end{equation}
We resort as before to a perturbative resolution for the
vertex, expressing the solution of 
(\ref{selfcons}) as a power series in the coupling 
$\lambda_0 = e_0^2/4\pi v_F $
\begin{equation}
\Gamma_3 ({\bf 0},0;{\bf k},\omega_k) = 
 1 + \sum_{n=1}^{\infty} \lambda_0^n 
                           \frac{s_n }{|{\bf k}|^{n\epsilon}} 
\label{ser}
\end{equation}
Successive terms in (\ref{ser}) can be obtained by means of an 
iterative procedure, using the formula
\begin{equation}
 \frac{e^2_0}{4} \int \frac{d^D p}{(2\pi )^D} \frac{1}{|{\bf p}|^{n\epsilon} }
            \frac{1}{v_F |{\bf p}|} \frac{1}{|{\bf k}-{\bf p}|} 
 =  \lambda_0  \frac{C_n(\epsilon)}{|{\bf k}|^{(n+1)\epsilon}}  
\label{rel3}
\end{equation}
where
\begin{equation}
C_n(\epsilon) = \frac{(4\pi )^{\epsilon /2}}{4}   
  \frac{\Gamma \left(\tfrac{n+1}{2} \epsilon  \right) \Gamma \left(\tfrac{1-(n+1)\epsilon}{2} \right) 
                                \Gamma \left(\tfrac{1-\epsilon}{2} \right) }
  { \sqrt{\pi} \Gamma \left(\tfrac{1+n\epsilon}{2} \right) \Gamma \left(1-\tfrac{n + 2}{2}\epsilon \right) }
\end{equation}
We get therefore the relation
\begin{equation}
s_{n+1} =  C_n(\epsilon) \: s_n
\label{rec3}
\end{equation}

It is remarkable that the recursion (\ref{rec3}) almost coincides with 
(\ref{rec}), the deviation being a term that vanishes in the limit 
$\epsilon \rightarrow 0$. We have in fact the result 
\begin{equation}
C_n(\epsilon) = \frac{D}{2} A_n(\epsilon) + \frac{1}{2} B_n(\epsilon)
\end{equation}
The agreement between the two recursion for $\Gamma_3 $ and 
$\widetilde{\Gamma}_{ij}$ would be perfect if we had a factor $D/2$ 
multiplying the right-hand-side of (\ref{re1}), which is 
instead missing due to the second-rank character of the tensor renormalized 
there. It can be shown that the precise consideration of this tensor character 
is a must for the correct implementation of a gauge-invariant renormalization, 
as observed for instance in the case of the electron current 
vertex\cite{jhep}. We are going to see that the mismatch in the two
ladder series, when combined with the effect of the electron self-energy 
corrections, leads to significant differences in the renormalization factors 
of the vertices $\Gamma_3$ and $\widetilde{\Gamma}_{ij}$.

It can be checked that a multiplicative renormalization by a factor $Z_m$ is 
enough to absorb the divergences in (\ref{ser}). Thus, we define 
the renormalized vertex, finite at $\epsilon \rightarrow 0$, by  
\begin{equation}
\Gamma_{3 , {\rm ren}} = Z_m \Gamma_3
\label{multh}
\end{equation}
The pole structure of the renormalization factor takes the form
\begin{equation}
Z_m = 1 + \sum_{i=1}^{\infty} \frac{d_i (\lambda )}{\epsilon^i}
\end{equation}

At this point, we can consider again the effect of the electron self-energy
corrections on the vertex by simply replacing the constant $v_F$ inside the 
integral in Eq. (\ref{selfcons}) by the effective Fermi velocity (\ref{veff}). 
After renormalizing $v_F$ by the factor $(\ref{zv})$, it becomes possible to 
absorb all the poles at $\epsilon = 0$ in the vertex by an appropriate choice 
of coefficients $d_i (\lambda )$ depending only on the renormalized coupling
$\lambda$. The first terms in the perturbative expansion can be computed 
analytically, with the result that 
\begin{eqnarray}
d_1 (\lambda )  & = &   - \frac{1}{2} \lambda - \frac{1}{8} \log(2) \: \lambda^2 
      - \frac{\pi ^2 + 120 \log ^2(2)}{1152}   \lambda^3               
        -  \frac{10 \pi ^2 \log (2)+688 \log ^3(2)+15 \zeta (3)}{6144}   \lambda^4   
                                                               \nonumber  \\
 & &  - \frac{13 \pi ^4+2064 \pi ^2 \log ^2(2)
        +144 \left(716 \log ^4(2)+37 \log (2) \zeta (3)\right)}{737280} \: \lambda^5   
                                                   +    \ldots   \label{firstd}     \\
d_2 (\lambda ) & = &  \frac{1}{16} \: \lambda^2 + 
            \frac{1}{24} \log(2) \: \lambda^3                         
         + \frac{5 \pi ^2 + 744 \log ^2(2)}{18432}  \lambda^4          \nonumber  \\      
 & &    + \frac{110 \pi ^2 \log (2)+8592 \log ^3(2)+135 \zeta (3)}{184320}  \: \lambda^5 
                                                            +  \ldots                     \\
d_3 (\lambda ) & = & - \frac{1}{768} \log (2) \: \lambda^4 
           - \frac{\pi ^2+360 \log ^2(2)}{184320} \lambda^5       + \ldots     \\
d_4 (\lambda ) & = & - \frac{1}{7680} \log (2) \: \lambda^5      +  \ldots  
\label{lastd}
\end{eqnarray}

The most important effect of the renormalization is again the anomalous 
scaling of the vertex $\Gamma_{3 , {\rm ren}}$, arising from its dependence 
on the momentum scale $\mu $ \cite{amit},
\begin{equation}
\Gamma_{3, {\rm ren}} \sim  \mu^{\gamma_m}
\end{equation}
The anomalous dimension $\gamma_m$ can be obtained from the renormalization 
factor as
\begin{equation}
\gamma_m = \frac{\mu }{Z_m}  \frac{\partial Z_m}{\partial \mu}
\end{equation}
This equation turns out to have a finite limit $\epsilon \rightarrow 0$
provided the set of conditions
\begin{equation}
\frac{d}{d\lambda } d_{i+1} (\lambda)
   - d_i (\lambda) \: \frac{d}{d\lambda } d_1 (\lambda)
   + b_1 (\lambda) \: \frac{d}{d\lambda } d_i (\lambda) = 0
\label{cond}
\end{equation}
are fulfilled\cite{ram}. In that case, the anomalous exponent is given by
\begin{equation}
\gamma_m = -\lambda \frac{d}{d\lambda } d_1 (\lambda)
\label{gamm}
\end{equation}
therefore depending solely on the renormalized coupling $\lambda $.

Quite nicely, it can be seen that the analytic expressions in 
(\ref{firstd})-(\ref{lastd}) satisfy the equations (\ref{cond}). We have also 
checked by numerical computation that the perturbative expansions of the 
functions $d_i (\lambda)$ fulfill that set of conditions, at least up to the 
order $\lambda^{24}$ we carried out the calculation. We can therefore 
rely on Eq. (\ref{gamm}) to obtain a sensible result for the anomalous 
dimension $\gamma_m$. The important point is that, as well as in the case of 
$\gamma_h$, the power series in $\lambda $ turns out to have a finite radius 
of convergence. This can be already noticed in the representation of the first 
orders of the expansion for $d_1 (\lambda)$, plotted in Fig. \ref{two}. The 
accurate fit achieved in Ref. \cite{jhep} (from the expansion to order 
$\lambda^{24}$) allows to get the value for the critical coupling 
\begin{equation}
\lambda_c' \approx 0.5448
\label{crit2}
\end{equation}

The critical coupling $\lambda_c'$ should mark the onset of a new phase in 
the electron system, characterized by a nonvanishing expectation value of the 
operator $\rho_3 ({\bf r})$. We observe however that the critical value 
(\ref{crit2}) is larger than the result (\ref{crit}) obtained for the coupling 
$\lambda_c$. This implies that the susceptibility to the development of loop 
currents should be stronger than that for the formation of a staggered charge 
density in graphene. Consequently, the phase with condensation of circular 
currents should take place first, as long as the system has a relative 
interaction strength above the critical coupling given by (\ref{crit}).

\section{Discussion}

We have seen that, at sufficiently large interaction strength, it is possible
to dynamically generate a mass term breaking parity (the so-called Haldane 
mass) in the many-body theory of Dirac fermions describing the graphene layer.
We have shown that the critical coupling for the development of such a mass
is below that corresponding to the usual excitonic instability, so that the 
dynamical breakdown of parity should prevail in graphene over the opening of 
a gap in the electronic spectrum from chiral symmetry breaking. 

As already mentioned, the mismatch between the critical couplings for the 
dynamical breakdown of parity and chiral symmetry may be paradoxical, given
that the commutation properties of the vertices for $\gamma_0$ and the current 
$i \epsilon_{3ij} \gamma_i \gamma_j $ are the same for Dirac fermions at
$D = 2$. We have clarified that the difference between the respective 
many-body corrections to the two vertices arises from the need to regularize 
high-energy contributions in a gauge-invariant manner, by analytic continuation 
of the momentum integrals below dimension $D = 2$. The suitability of this 
method to produce precise physical results for the carbon layer has been 
recently checked in Ref. \onlinecite{ros}. In this regard, the different 
strength of parity and chiral symmetry breaking is a quantum field theory  
^^ ^^ anomaly" arising as a remnant of high-energy contributions to the 
low-energy effective theory of graphene, similar to the
anomalous non-conservation of the chiral current in conventional Quantum 
Electrodynamics at $D = 3$ when a gauge invariant method is used to regularize
the theory.

\begin{figure}

\vspace{0.5cm}

\begin{center}
\mbox{\epsfxsize 7cm \epsfbox{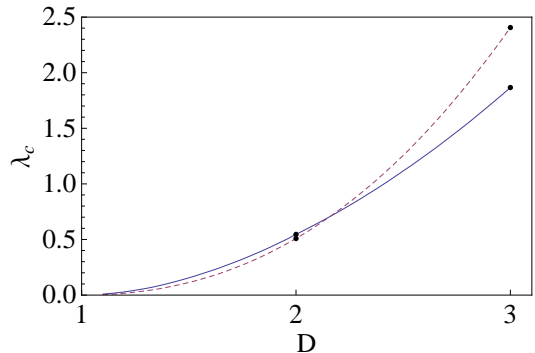}} 

\end{center}
\caption{Plot of the critical couplings $\lambda_c $ for the dynamical 
symmetry breaking of parity (points on the dashed line) and for the breakdown
of chiral symmetry (points on the full line) depending on the spatial
dimension $D$ of the interacting electron system.}
\label{three}
\end{figure}

It is interesting to observe that the above computation of anomalous exponents
can be also applied to investigate the dynamical symmetry breaking of
the many-body theory of Dirac fermions at any spatial dimension $D$. This is
so as the classical scale invariance (\ref{scale}) can be extended to hold
for the action of the Dirac fermions interacting with the same $1/|{\bf r}|$
Coulomb potential at arbitrary $D$. It can be shown then that all cutoff 
dependences can be always absorbed into a finite number of renormalization 
factors, at least in the ladder approximation supplemented with electron 
self-energy corrections. The anomalous exponents of the order parameters 
corresponding to $\gamma_0$ and the current $i \epsilon_{3ij} \gamma_i \gamma_j $ 
display in general finite radii of convergence in our ladder approach, which 
we have computed for spatial dimension $1 < D \le 3$. The results are shown 
in Fig. \ref{three}. We observe for instance that the critical coupling for 
the conventional excitonic instability is the lowest at $D = 3$, while we 
find the reverse situation for $D \le 2$. At $D = 2$, it is actually the 
need to regularize the many-body theory by computing in the limit 
$D = 2 - \epsilon $ which makes the dynamical breakdown of parity to prevail 
over chiral symmetry breaking at the integer physical dimension.

Regarding the actual value of the critical coupling (\ref{crit}), we may ask 
whether there is any chance to observe the effects of a parity-breaking 
mass in real graphene samples. At this point, we note that the coupling 
$\lambda_c$ cannot be compared directly to nominal values of the graphene
fine structure constant, as $\lambda $ should correspond in our approach to
the effective coupling after screening of the interaction. That is, the value 
of $\lambda_c$ must be referred to the relative strength of the dressed 
Coulomb potential. Under the assumption of static RPA screening of the 
interaction, we get for instance the relation between the effective coupling 
$\lambda $ and the nominal value $\alpha $ of the graphene fine 
structure constant
\begin{equation}
\lambda = \frac{\alpha }{1+ \frac{\pi}{2}\alpha }
\end{equation}
The above formula takes into account the effect of two fermion flavors with
different spin projection. In the case of graphene isolated in vacuum, the 
bare value $\alpha \approx 2.2$ leads to $\lambda \approx 0.49$, which is very 
close to the computed critical value (\ref{crit}). We recall anyhow that the 
use of the static RPA tends to underestimate appreciably the relative strength 
of the dressed interaction. The value of $\lambda $ obtained after dynamical 
screening of the interaction has been shown to be (starting from 
$\alpha \approx 2.2 $) above the critical coupling $\lambda_c' $ in 
(\ref{crit2}) \cite{ggg,prb}. Therefore, we are led to conclude that the 
electronic state of graphene in vacuum should correspond to the phase with 
dynamical breakdown of parity above $\lambda_c $.

We note finally that the phase with the dynamical breaking of parity may
not have necessarily insulating character, as it corresponds to the 
spontaneous development of an order parameter (a vector) with two possible 
orientations. We may think of situations in which the space is divided into 
different domains where the order parameter alternates between opposite 
directions. In these configurations, electronic transport may be possible 
along the interface between neighboring domains. The study of the dynamics of 
this kind of disorder at a microscopic level is beyond the scope of this work, 
but it is plausible that, for suitably small domains, the phase with the 
dynamical breaking of parity may still have metallic properties. The 
development of an ordered phase over macroscopic regions may require some 
external source contributing to the complete alignment of the order parameter. 
This can be achieved for instance by introducing a transverse magnetic field, 
setting in this way a preferred orientation in the parity-breaking mechanism. 
Signatures of the onset of the ordered phase may have been already seen in the 
metal-insulator transition observed in graphene at relatively low magnetic 
fields\cite{check,zhang}. Further efforts should be devoted to the detection 
of the broken-symmetry phase, which should anyhow appear in the close 
proximity (about 1 meV or below) around the charge neutrality point.

\section{Acknowledgments}

We thank F. de Juan, F. Guinea,  V. Juri\v{c}i\'c and C. P. Mart\'{\i}n 
for useful insights into the subject. 
The financial support from MICINN (Spain) through grants FIS2008-00124/FIS 
and FIS2011-23713 is gratefully acknowledged.



\end{document}